\documentclass[onecolumn, floatfix, preprintnumbers, eqsecnum, letterpaper, superscriptaddress, nofootinbib]{revtex4}
\usepackage{graphicx}
\usepackage{microtype}
\usepackage{amsmath}
\usepackage{amssymb}
\usepackage{subfigure}
\usepackage{hyperref}
\usepackage{url}
\usepackage{xcolor}
\usepackage{color}
\usepackage{mathrsfs}
\usepackage{calrsfs}
\usepackage{amsfonts}
\usepackage{eufrak}
\usepackage{tabularx}
\usepackage{eucal}
\usepackage{latexsym}
\usepackage{ragged2e}
\usepackage{epsfig}
\usepackage{textcomp}
\usepackage{caption}
\usepackage{subfigure}
\usepackage{marvosym}
\usepackage{ifsym}
\usepackage{lipsum}

\DeclareCaptionJustification{justified}{\leftskip=0pt \rightskip=0pt \parfillskip=0pt plus 1fil}
\definecolor{vividviolet}{rgb}{0.62, 0.0, 1.0}
\definecolor{amaranth}{rgb}{0.9, 0.17, 0.31}
\definecolor{palatinateblue}{rgb}{0.15, 0.23, 0.89}
\definecolor{brightpink}{rgb}{1.0, 0.0, 0.5}
\definecolor{cornflowerblue}{rgb}{0.39, 0.58, 0.93}
\definecolor{deepcarminepink}{rgb}{0.94, 0.19, 0.22}
\definecolor{radicalred}{rgb}{1.0, 0.21, 0.37}

\hypersetup{ linktoc=all,
    colorlinks, linkcolor={palatinateblue},
    citecolor={brightpink}, urlcolor={amaranth}
}

\graphicspath{{Images/}}

\graphicspath{{Images/}}

\def\@fnsymbol#1{\ensuremath{\ifcase#1\or \ddagger \or  $\textleaf$  \or \dagger
\else\@ctrerr\fi}}%

\makeatother

\def\sideremark#1{\ifvmode\leavevmode\fi\vadjust{\vbox to0pt{\vss
 \hbox to 0pt{\hskip\hsize\hskip1em
 \vbox{\hsize1.3cm\tiny\raggedright\pretolerance10000
 \noindent #1\hfill}\hss}\vbox to8pt{\vfil}\vss}}}%

\def\beq{\begin{equation}}
\def\eeq{\end{equation}}

 \newcommand{\be}{\begin{equation}}
	\newcommand{\en}{\end{equation}}

\begin{document}

\title{\Large First Principle Study of Gravitational Pressure and Thermodynamics\\
 \vspace{0.2cm}
 of FRW Universe}

\author{Haximjan Abdusattar}
\email{axim@nuaa.edu.cn}
\affiliation{College of Physics, Nanjing University of Aeronautics and Astronautics, Nanjing, 211106, China}

\author{Shi-Bei Kong}
\email{kongshibei@nuaa.edu.cn}
\affiliation{College of Physics, Nanjing University of Aeronautics and Astronautics, Nanjing, 211106, China}

\author{Wen-Long You}
\email{wlyou@nuaa.edu.cn}
\affiliation{College of Physics, Nanjing University of Aeronautics and Astronautics, Nanjing, 211106, China}
\affiliation{Key Laboratory of Aerospace Information Materials and Physics (NUAA), MIIT, Nanjing 211106, China}

\author{Hongsheng Zhang }
\email{sps\_zhanghs@ujn.edu.cn}
\affiliation{School of Physics and Technology, University of Jinan, 336 West Road of Nan Xinzhuang, Jinan, Shandong 250022, China}
\affiliation{Key Laboratory of Theoretical Physics, Institute of Theoretical Physics, Chinese Academy of Sciences, Beijing 100190, China}

\author{Ya-Peng Hu \Letter}
\email{huyp@nuaa.edu.cn}
\affiliation{College of Physics, Nanjing University of Aeronautics and Astronautics, Nanjing, 211106, China}
\affiliation{Key Laboratory of Aerospace Information Materials and Physics (NUAA), MIIT, Nanjing 211106, China}
\affiliation{Center for Gravitation and Cosmology, College of Physical Science and Technology, Yangzhou University, Yangzhou 225009, China}

{\let\thefootnote\relax\footnotetext{\vspace*{0.2cm} \Letter Corresponding author}}

\begin{abstract}

 We make a first principle study of gravitational pressure in cosmic thermodynamics. The pressure is directly derived from the unified first law, in fact the Einstein field equation in spherically symmetric spacetime. By using this pressure, we obtain the thermodynamics for the FRW universe, especially presenting the gravitational equation of state for the FRW spacetime itself, i.e. $P=P(R_A, T)$ for the first time. Furthermore, we study the Joule-Thomson expansion as an application of the thermodynamic equation of state to find the cooling-heating property of the FRW universe. We demonstrate that there is an inversion temperature for a FRW universe if its enthalpy ${\cal H}$ is negative. These investigations shed insights on the evolution of our universe in view of thermodynamics.

\end{abstract}

\maketitle

\section{Introduction}

Black hole thermodynamics takes a pivotal status in modern theoretical physics. It is an arena of thermodynamics, statistical mechanics, gravity and quantum mechanics. In particular the entropy of black hole presents a sharpest probe into quantum gravity \cite{Hawking:1974sw,Bardeen1973,Bekenstein:1974ax}. Nowadays black hole thermodynamics already goes beyond the study of black hole itself. The thermodynamic concepts such as gravitational entropy and temperature, have been applied to other gravity-controlled systems, e.g.\ FRW universe \cite{Cai:2005ra,Gong:2007md} and pure dS spacetime \cite{Gibbons:1977mu}, where no black hole appears.
In all these systems, significant research progresses have been made in thermodynamics, which implies essential relationship between gravity and thermodynamics.

Traditional black hole thermodynamics suffers from an intractable problem in applications to realistic celestial objects, that is, the concept of event horizon is teleological, i.e.\ one has to acquire all the information at future infinity to determine the present thermodynamic quantities. An interesting approach to solve this problem is the unified first law, in which only quasi-local quantities are involved. For spherically symmetric spacetime, the unified first law can be derived from the Einstein field equation.
In spherically symmetric spacetime, one can define a conserved charge, which is the Misner-Sharp energy \cite{Hayward:1997jp,Cai:2006rs,Maeda:2007uu,Cai:2009qf,Hu:2015xva},
\begin{equation}
 M_{_{MS}}=-\frac{1}{8\pi} \int *(T_{ab}K^{b})\,,
\end{equation}
for the Kodama vector $K^a$ corresponding to the conserved current $T_{ab}K^{b}$. By using this conserved charge, one obtains the general form of the unified first law,
\begin{eqnarray} \label{UFdE0}
d M_{MS}=\widetilde{A} \widetilde{\Psi}_{i}dx^{i}+\widetilde{W} d \widetilde{V}
\end{eqnarray}
(see \ref{sII} for detailed explanation). In the above equation, all the quantities are quasi-locally measured, and the last term is related to a generalized force.

In this quasi-local progress of gravitational thermodynamics, one only needs to consider apparent horizon, and thus the teleological problem is evaded. Very importantly a generalized force term appears naturally, which is identified as the spacetime pressure. In some studies of AdS black hole thermodynamics, pressure was also invoked to trigger phase transition in extended phase space \cite{Kubiznak:2012wp,Xu:2015rfa,Kubiznak:2016qmn,Hu:2018qsy} (see also  \cite{Gunasekaran:2012dq,Wei:2012ui,Hendi:2012um,Cai:2013qga,Altamirano:2013ane,Bhattacharya:2017nru,Li:2020xkh,Hu:2020pmr} for related works). However, pressure in these studies is only an analogy of the effects of the cosmological constant in cosmology $P_\Lambda=-{\Lambda}/{8\pi}$ \cite{Kastor:2009wy,Dolan:2010ha,Cvetic:2010jb,Debnath:2020inx}. A noteworthy point in the phase transition in extended phase space is that different cosmological constants correspond to different gravity theories. Thus the extended phase space represents the phase space of different theories rather than different solutions of a theory. This is a quite interesting progress from the viewpoint of ensemble theory. We will study the thermodynamics of FRW universe from the unified first law, and thus directly from Einstein field equation. One will see that pressure gets a natural definition.

Friedmann-Robertson-Walker (FRW) universe is a dynamical space-time. It has been widely accepted that the FRW universe has thermodynamics embodied on the apparent horizon, whose area plays the role of the entropy and surface gravity behaves as temperature. In spirit of Jacobson's derivation \cite{Jacobson:1995ab} of the Einstein field equations from the Clausius relation, Cai and Kim \cite{Cai:2005ra} first investigated the Friedmann equations of the FRW universe on the apparent horizon, whose thermodynamics is associated with the unified first law \cite{Cai:2006rs,Cai:2008ys}. A similar connection between the Friedmann equations and the first law of thermodynamics has also been discovered in alternative theories of gravity \cite{Akbar:2006kj,Akbar:2006mq}. Related approaches on black hole thermodynamics can be found in \cite{Zhang:2014ala}. Researches on the thermodynamics of FRW universe shed light on inherent relationship between thermodynamics and gravity. It may be a universal and fundamental finding that would be helpful to decode the nature and the quantization of gravity.

To our knowledge, the thermodynamics of the FRW universe has not been deeply and extensively investigated. Besides the laws of thermodynamics, to describe the behavior of a thermodynamic system, it is usually crucial to construct its thermodynamic equation of state. This is difficult for a FRW universe, unless we find a proper definition of its thermodynamic pressure corresponding to the physics related to the apparent horizon.
It should be emphasized that, different from Refs.\ \cite{Kastor:2009wy,Dolan:2010ha,Cvetic:2010jb,Debnath:2020inx} where the pressure is simply defined by the cosmological constant,
in this paper we discover a more suitable definition of the thermodynamic pressure
for the FRW universe
\begin{equation}\label{P}
P\equiv W-P_{\Lambda},
\end{equation}
where $W$ is the work density of the matter field $W:=-h_{i j}T^{i j}/2$, and $P_{\Lambda}=-\Lambda/8\pi$ is the conventional pressure defined for the AdS black hole thermodynamics. In addition, $h_{i j}$ and $T^{i j}$ are the $0,1$-components of the metric and the stress-tensor \cite{Akbar:2006kj,Kong:2021qiu} with $i,j=0,1,\quad x^0=t, x^1=r$. By using this new definition of pressure, we derive the thermodynamic equation of state $P = P(R_A, T)$ for a FRW spacetime with the cosmological constant for the first time.

Furthermore, we will study Joule-Thomson expansion of the FRW universe as an application of the thermodynamic equation of state $P = P(R_A, T)$. In classical thermodynamics, the Joule-Thomson expansion describes the expansion of gas passing through a throat from a high pressure region to a low pressure region with a constant enthalpy. In the framework of extended black hole thermodynamics, Ref.\cite{Okcu:2016tgt} creatively applied the famous Joule-Thomson expansion of the van der Waals systems to the charged AdS black hole, and found the existence of the inversion temperature and inversion pressure. Since then, a great deal of attention has been drawn to this topic in various black holes system, such as Kerr\textendash{}AdS black holes \cite{Okcu:2017qgo}, Gauss-Bonnet black holes \cite{Lan:2018nnp}, regular (Bardeen)-AdS black holes~\cite{Pu:2019bxf,Li:2019jcd,Rajani:2020mdw}, and Born-Infeld AdS black holes \cite{Bi:2020vcg}, see more in Refs.\cite{Ghaffarnejad:2018exz,MahdavianYekta:2019dwf,Mo:2018rgq,Kuang:2018goo,Cisterna:2018jqg,Liang:2021xny}. It would be interesting, and non-trivial to find the corresponding results in the thermodynamics of the universe. In this work, we will discuss the Joule-Thomson expansion of the FRW universe in analogy to AdS black holes. We demonstrate that the existence of inversion temperature and inversion pressure depends on the negative value of enthalpy ${\cal H}$ for a FRW universe, i.e.~${\cal H}<0$.

The organization of this paper is as follows. In Sec.\ref{sII}, we review thermodynamics of FRW universe on the apparent horizon in Einstein gravity without
cosmological constant. In Sec.\ref{sIII}, we study the unified first law for a FRW universe with the cosmological constant, and further derive the thermodynamic equation of state $P = P(R_A, T)$. In Sec.\ref{sIV}, we study the Joule-Thomson expansion of FRW universe as an application of the thermodynamic equation of state. In Sec.\ref{sV}, we present conclusions and discussion of this paper.

\section{Review: Previous results of the thermodynamics in a FRW Universe}\label{sII}

In this section, for convenience to make our main results more readable, we will briefly review the thermodynamics in a FRW Universe with zero cosmological constant in previous work, i.e. the apparent horizon, Hawking temperature and the first law of thermodynamics for a FRW universe in Einstein gravity with the absence of cosmological constant. In an isotropic coordinate system $x^\mu=(t, r, \theta, \varphi)$, the line element of a FRW universe is written
\begin{equation}\label{FRW}
ds^2=-dt^2 +a^2(t)\left[\frac{dr^2}{1-kr^2}+r^2\left(d\theta^2 + \sin^2 \theta d\varphi^2 \right)\right]\,,
\end{equation}
where $a(t)$ is a scale factor describing the evolution of the universe, and $k=+1, 0, -1$ are the spatial curvatures corresponding to a spherical, flat and hyperbolic universes,
respectively\footnote{The present observations imply that $\Omega_k\sim (-0.02, 0) $ (depending on different observations and different joint analysis) \cite{DiValentino:2019qzk,Handley:2019tkm,DiDio:2016ykq,Bel:2022iuf}. In fact, one never fixes the spatial curvature to be exactly zero from observations. Theoretically, the spatial curvature is not an objective quantity in relativity theory, which depends on sling or observers.}. If one introduces the areal radius $R(t,r)\equiv a(t)r$, the line element (\ref{FRW}) can be further rewritten as
\begin{equation} \label{nFRW}
ds^2=h_{ij}dx^{i} dx^{j}+R^2(t,r)\left(d\theta^2 + \sin^2 \theta d\varphi^2 \right)\,,
\end{equation}
where $x^{0}=t$, $x^{1}=r$, $h_{ij}={\rm diag}\Big(-1,~\frac{a^2(t)}{1-kr^{2}}\Big)$. Note that, this metric is convenient to investigate the thermodynamics for a dynamical spacetime with the spherical symmetry, and for simplicity we denote $a=a(t)$, $R=R(t,r)$ in the following.

The apparent horizon is a trapped surface with vanishing expansion $h^{i j}\partial _{i}R\partial _{j}R=0$ \cite{Hayward:1994bu}, and hence the apparent horizon of a FRW universe $R_{A}$ is easily solved as
\cite{Cai:2005ra}
\begin{equation}\label{AH}
R_{A}=\frac{1}{\sqrt{H^{2}+\frac{k}{a^{2}}}} \, ,
\end{equation}
where $H=H(t):=\dot a/a$ is the Hubble parameter characterizing the expansion rate of the universe, and it is clear that different $k$ corresponds to different size of the horizon with same the Hubble parameter. From the location of apparent horizon, we easily obtain a useful equality in the following as
\begin{equation}\label{dAH}
\dot{R}_{A}=-R_{A}^{3}H\Big(\dot{H}-\frac{k}{a^{2}}\Big) \, ,
\end{equation}
where the derivative of $R_{A}$ is with respect to the cosmic time $t$, and it describes the nature of time dependence of the apparent horizon.

The surface gravity on the apparent horizon is given by \cite{Cai:2006rs,Akbar:2006kj}
\begin{equation}\label{kappa}
\kappa=\frac{1}{2\sqrt{-h}}{\partial _i}\left(
\sqrt{-h}~h^{ij}{\partial _{j}R}\right) \, ,
\end{equation}
where $h={\rm det}(h_{ij})$. Therefore, the surface gravity of the apparent horizon in a FRW universe is derived as
\begin{equation}
\kappa=-\frac{1}{R_{A}}\Big(1-\frac{\dot{R}_{A}}{2H R_{A}}
\Big) \, , \label{sg}
\end{equation}
and hence the Hawking temperature on the apparent horizon of the FRW universe is
\begin{equation}\label{T1}
T\equiv\frac{|\kappa|}{2\pi}=\frac{1}{2\pi {R_A}}\Big(1-\frac{\dot{R}_A}{2H R_{A}}
\Big) \, ,
\end{equation}
where the surface gravity of the apparent horizon is usually negative in a FRW universe~\cite{Akbar:2006kj}, i.e. $\kappa<0$, which will impose a constraint on the matter field such as a perfect fluid, see Appendix \ref{appA} for related discussion. When the apparent horizon $R_{A}$ changes very slowly, i.e. ${\dot{R}_{A}}/(2 HR_{A})\ll 1$, the Hawking temperature is written simply as \cite{Cai:2008gw}
\begin{equation}\label{T2}
T\approx \frac{1}{2\pi R_{A}}\,.
\end{equation}
However, $\dot{R}_{A}$ is in fact not necessarily a small quantity, and hence we will use the general expression of temperature in Eq.(\ref{T1}) to perform the following investigations.

The first law of thermodynamics for a FRW universe is usually inspired from the well-known unified first law~\cite{Hayward:1997jp,Cai:2006rs,Hu:2015xva}
\begin{eqnarray} \label{UFdE}
d \widetilde {E}=\widetilde {A} \widetilde {\Psi}_{i}dx^{i}+\widetilde {W} d \widetilde {V}\,,
\end{eqnarray}
where $\widetilde {A}=4\pi R^2$ and $\widetilde {V}={4\pi R^3}/{3}$ are area and volume of the $3$-dimensional sphere with radius $R$. The work density $\widetilde {W}$ and energy-supply $\widetilde {\Psi}_{i}$ are defined as
\begin{equation}
\widetilde {W}:=-\frac{1}{2}h^{i j}T_{i j} ,~~~~~~~ \widetilde {\Psi}_{i}:=T_{i}^j{\partial _{j}R}+\widetilde {W}{\partial _{i}R}\,,
\end{equation}
where $T_{i j}$ is the projection of the energy-momentum tensor $T_{\mu \nu}$ of the matter in the $(t,r)$ directions. In addition, $\widetilde {E}$ is the Misner-Sharp energy inside the radius $R$ in a spherically symmetric spacetime with metric Eq.(\ref{nFRW}). In the Einstein gravity without cosmological constant, the
Misner-Sharp energy is given by \cite{Cai:2005ra,Cai:2006rs,Gong:2007md}
\begin{equation} \label{MSE}
\widetilde {E}=\frac{R}{2}(1-h^{i j}\partial _{i}R\partial _{j}R)\,.
\end{equation}
For a FRW filled with the perfect fluid
\begin{equation}\label{Tmu}
 T_{\mu \nu}=(\rho+p)u_{\mu}u_{\nu}+pg_{\mu \nu}\,,
\end{equation}
where the four-velocity of the fluid is $u^\mu=\frac{1}{\sqrt{-g_{tt}}}\delta_t^\mu$, $\rho$ and $p$ are the energy density and pressure,
respectively, one can further obtain
\begin{equation}
\widetilde {W}=-\frac{1}{2}(T^{t}_{t}+T^{r}_{r})=\frac{1}{2}(\rho-p) , ~~~~\widetilde {\Psi}\equiv\widetilde {\Psi}_{i}dx^{i}=\frac{1}{2}(\rho+p)(-HRdt+adr)\,.
\end{equation}
Therefore, after projecting the unified first law Eq.(\ref{UFdE}) on the apparent horizon of a FRW universe, one finally obtains the first law of thermodynamics for a FRW universe
\begin{eqnarray} \label{UFAH}
dE=\frac{\kappa}{8\pi}dA+W dV=-TdS+WdV\,,
\end{eqnarray}
where the expression of the temperature is in (\ref{T1}) and
\begin{eqnarray}
	E=\widetilde {E}|_{R=R_A}=\frac{R_A}{2}\ , \ \ S=\frac{\widetilde {A}}{4}|_{R=R_A} = \pi R_{A}^2 \ , \ \
    V=\widetilde {V}|_{R=R_A} =\frac{4}{3}\pi R_A^3 \ , \ \ W=\widetilde {W}|_{R=R_A}=\frac{1}{2}(\rho-p) \,.\label{MSVWdef}
\end{eqnarray}
It should be pointed out that the minus sign before $TdS$ in (\ref{UFAH}) arises from the surface gravity $\kappa$ on apparent horizon is negative, and hence one keeps the corresponding temperature $T$ positive \cite{Akbar:2006kj,Dolan:2013ft}. In addition, the minus sign also relates to the fact that the total energy flows into the apparent horizon.

\section{Thermodynamic Equation of State for a FRW universe with the Cosmological constant}\label{sIII}

In the previous section, the thermodynamics of a FRW universe is investigated in the absence of a cosmological constant. In this section, we derive the thermodynamic equation of state for a FRW universe with the cosmological constant, since the thermodynamic pressure in the AdS black hole cases has been found to relate with the cosmological constant. We first obtain the unified first law of thermodynamics by using the Misner-Sharp energy and Einstein field equations of a FRW universe with the cosmological constant.

Note that the original form (\ref{MSE}) for the Misner-Sharp energy is applicable for Einstein gravity without cosmological constant in four dimensions. In addition, a
generalization of the Misner-Sharp energy in presence of the cosmological constant $\Lambda$ has also been considered in Einstein gravity as well as the Einstein-Gauss-Bonnet
gravity, $f(R)$ gravity, etc.
In FRW universe, the Misner-Sharp energy with $\Lambda$ in Einstein gravity $\widetilde {M}$ is obtained from  \cite{Maeda:2007uu,Cai:2009qf,Hu:2015xva,Zhang:2019oes}
\begin{equation} \label{FME0}
\widetilde M=\frac{R}{2}\left[-\frac{\Lambda}{3}{R^2}+{R^2}\left(H^2 +\frac{k}{a^2}\right)\right] \, .
\end{equation}
The Einstein field equation with the cosmological constant is
\begin{equation}\label{Eeq}
  G_{\mu \nu}+\Lambda g_{\mu \nu} =8\pi T_{\mu \nu} \,.
\end{equation}
For the FRW metric (\ref{nFRW}) and energy-momentum tensor (\ref{Tmu}), the $(t,t)$ component of
the above Einstein field equation is easily written as
\begin{eqnarray}\label{FE}
H^2+\frac{k}{a^2}=\frac{8\pi}{3}\rho+\frac{\Lambda}{3}\,,
\end{eqnarray}
which is just the first Friedmann equation.
On the other hand, from the energy-momentum conservation law of matter fields $\nabla_{\nu}T^{\mu\nu}=0$, one finds the continuity equation~\cite{Zhang:2019oes}
\begin{eqnarray}\label{drho}
\dot{\rho}+3H(\rho+p)=0\,.
\end{eqnarray}
By differentiation of the Misner-Sharp energy (\ref{FME0}) and combination with Eqs. (\ref{UFdE}), (\ref{FE}) and (\ref{drho}), we have
\begin{eqnarray} \label{dM}
d \widetilde M&=& R^{3}H\left(\dot{H} -\frac{k}{a^2}\right)dt+\frac{R^2}{2}\left[3\left(H^2+\frac{k}{a^2}\right)-\Lambda \right]dR-\frac{R^3}{6}d{\Lambda} \nonumber\\
&=& -4 \pi R^3 H(\rho+p)dt +4\pi R^2 \rho dR-\frac{R^3}{6}d{\Lambda} \nonumber\\
&=& \widetilde A \widetilde \Psi+\widetilde W d \widetilde V -\frac{R^3}{6}d{\Lambda}\,,
\end{eqnarray}
which is just the unified first law of thermodynamics of a FRW universe with the cosmological constant $\Lambda$. Here, similar with studies of critical phenomenon in AdS black hole thermodynamic cases with the cosmological constant, we consider the cosmological constant in Eq.(\ref{dM}) as a variable quantity\footnote{Different cosmological constants corresponds to different FRW universes in an ensemble. This method is taken in many modern theories, e.g. the landscape of string theory.},
and identify the cosmological constant as a pressure
\begin{equation}\label{PLambda}
P_{\Lambda}=-\frac{\Lambda}{8\pi}\,.
\end{equation}
Then the Eq.(\ref{dM}) can be rewritten as
\begin{eqnarray}\label{MH}
d \widetilde M=\widetilde A \widetilde \Psi+\widetilde W d \widetilde V+\widetilde V d P_{\Lambda}\,.
\end{eqnarray}
From which, we can find that the volume $\widetilde V={4\pi R^3}/{3}$ of $3$-dimensional sphere with radius $R$ conjugates to the pressure $P_{\Lambda}$, too. At the apparent horizon, similar with Eq.(\ref{UFAH}), one gets the first law of thermodynamics in the extended phase space of a FRW universe with the cosmological constant
\begin{eqnarray} \label{UFAH1}
dM=\frac{\kappa}{8\pi}dA+W dV+VdP_{\Lambda}=-TdS+W dV+VdP_{\Lambda}\,,
\end{eqnarray}
where
\begin{eqnarray}
M=\widetilde {M}|_{R=R_A}=\frac{R_A}{2}(1-\frac{\Lambda}{3}{R_A^2}),~ S=\frac{\widetilde {A}}{4}|_{R=R_A} = \pi R_{A}^2,~
V=\widetilde {V}|_{R=R_A} =\frac{4}{3}\pi R_A^3,~ W=\widetilde {W}|_{R=R_A}=\frac{1}{2}(\rho-p)\,.\label{MSVWdef1}
\end{eqnarray}

It should be emphasized that there are two differences from studies of critical phenomenon in AdS black hole thermodynamic cases with the cosmological constant. One is the minus sign before $TdS$, while the other is that an extra work term $WdV$ is added. Since $W$ can be also considered as the pressure, a subtlety of choosing a suitable pressure to obtain the thermodynamic equation of state for a FRW universe with the cosmological constant appears. In our this paper, we will follow the spirit of a proper definition from the first law of thermodynamics to find a suitable thermodynamic pressure. From Eq.(\ref{UFAH1}), in fact we can further rewrite it as the standard form
\begin{eqnarray} \label{UFAH2}
dU=TdS-PdV\,,
\end{eqnarray}
where the corresponding internal energy $U$ and pressure $P$ are \footnote{\label{ftn3} Note that, we have put an extra minus sign in front of $M$ in (\ref{UP}) to let $U$ absorb the minus sign in front of the $TdS$-term in the first law (\ref{UFAH1}) (see e.g. \cite{Banihashemi:2022jys,Banihashemi:2022htw} for various ways of treating this minus sign).
This treatment results in a negative internal energy $U=-R_A/2<0$, which appears to be a little strange. In fact, whether the internal energy is negative depends on the choice of the reference state. In principle one could alternatively define the internal energy as $\widetilde U=-M+P_\Lambda V+U_0=U+U_0$, where $U_0$ is a constant that can be fixed by the choice of the reference state, and hence $\widetilde U$ is not necessarily negative. In our paper, we have chosen the reference state at $R_A=0$, that is, we have set $\widetilde U(R_A=0)=0$ and hence $U_0=0$. However, here we should emphasize that it is $d\widetilde U=dU$ instead of $\widetilde U$ that is of physical importance in the context of our paper. This can be easily seen from the fact that the choice of $U_0$ does not affect the relation shown in the first law (\ref{UFAH2}). More importantly, in the following part of this paper, since the definition of thermodynamic pressure $P$ is directly read off from the first law, the equation of state will be also independent of the choice of $U_0$.}
\begin{equation}\label{UP}
U\equiv-M+VP_{\Lambda}\,, ~~~~P\equiv W-P_{\Lambda}\,.
\end{equation}
Obviously, here the pressure $P$ will be more suitable to be the thermodynamic pressure for a FRW universe with the cosmological constant, and volume $V=\frac{4}{3}\pi R_A^3$ is just the thermodynamic volume conjugates to thermodynamic pressure $P$.

In the following, basing on the above proper definition of thermodynamic pressure $P$, we derive the thermodynamic equation of state in the extended phase space for a FRW universe with the cosmological constant. Regarding the apparent horizon (\ref{AH}) and from (\ref{FE}), one further
gets
\begin{equation}\label{rho}
\rho=\frac{3}{8\pi R_{A}^{2}}-\frac{\Lambda}{8\pi}\,.
\end{equation}
From (\ref{dAH}), (\ref{drho}) and (\ref{rho}), $p$ is rewritten as
\begin{equation}\label{p}
p=-\frac{3}{8\pi R_{A}^{2}}+\frac{\Lambda}{8\pi}+\frac{\dot{R}_{A}}{4\pi H R_{A}^{3}}\,.
\end{equation}
Note that the variation rate of the apparent horizon in (\ref{p}) can be replaced into the temperature $T$ from solving (\ref{T1}). Then, from (\ref{rho}) and (\ref{p}) combined with (\ref{UP}), we easily obtain the thermodynamic equation of state as\footnote{Note that, the cosmological constant $\Lambda$ and the curvature parameter $k$ do not explicitly appear in the equation of state. In fact, $\Lambda$ and $k$ affect the values of $P$, $T$ and $R_A$ but do not affect the relation among these values.}
\begin{eqnarray}\label{PVT}
P=\frac{T}{2R_{A}}+\frac{1}{8 \pi R_{A}^2}\,.
\end{eqnarray}
For Eq.(\ref{PVT}), we can also easily rewrite it in physical units according to the dimensional analysis
\begin{eqnarray}\label{PVTchkG}
\texttt{P}=\frac{k_B\texttt{T}}{2\ell_p^2 R_{A}}+\frac{\hbar c}{8\pi \ell_p^2 R_{A}^2}\,,
\end{eqnarray}
where the physical pressure $\texttt{P}$ and temperature $\texttt{T}$ in physical units are related to $P$ and $T$ as
\begin{equation}\label{physical PT}
  \texttt{P}=\frac{\hbar c}{\ell_p^2}P \,,~~~~~~~~~~~ \texttt{T}= \frac{\hbar c}{k_B}T \,,
\end{equation}
and the square of the Planck length is $\ell_p^2=\hbar G/c^3$, while $k_B$ is the Boltzmann constant. In order to have
an insightful view of the equation of state (\ref{PVTchkG}), some isothermal curves from it are drawn in the $P$-$R_A$ plane, see the following Fig.\ref{Fig1}.

 \begin{figure}[h]
 \begin{minipage}[t]{1\linewidth}
\includegraphics[width=8cm]{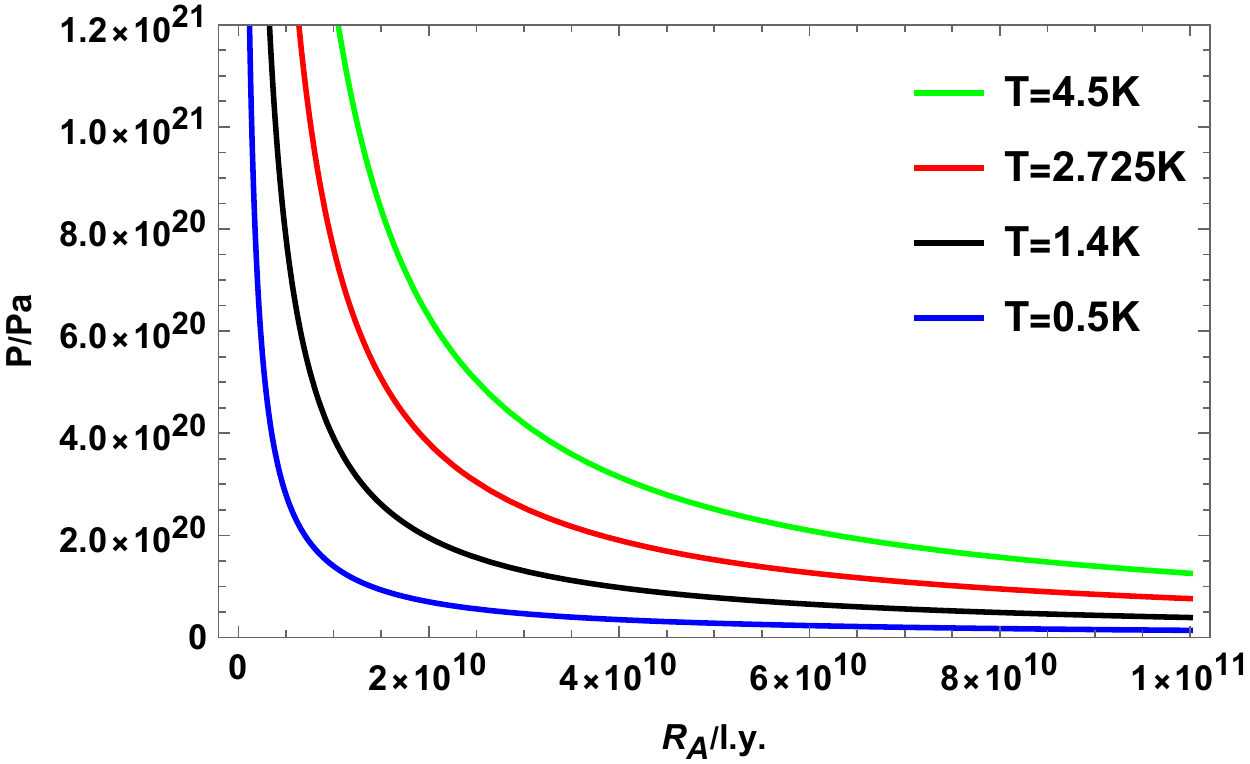}
 \end{minipage}
\caption{\footnotesize
Isothermal curves of FRW universe in the $P$-$R_A$ phase diagram, which imply no $P$-$V$ phase transition.
}
\label{Fig1}
\end{figure}
In Fig.\ref{Fig1}, we plot some numerical results of the pressure $P$ as function of the apparent horizon $R_{A}$ with some fixed values of $T$. The values of the temperatures are chosen such that the interesting features can be nicely plotted. Our choice is therefore for convenience only.
From the figure, there is no evidence for the existence of van der Waals-like ($P$-$V$) phase transition, which can be also rigorously proved in the following. The necessary condition for the existence of the $P$-$V$ phase transition
\cite{Kubiznak:2012wp,Hu:2018qsy,Wei:2012ui,Gunasekaran:2012dq,Cai:2013qga}
 is
\begin{equation}\label{PV}
\left(\frac{\partial P}{\partial V}\right)_{T}=\left(\frac{\partial^2 P}{\partial V^2}\right)_{T}=0\,,
\end{equation}
or equivalently in this paper
\begin{equation}\label{PVTc}
\left(\frac{\partial P}{\partial R_A}\right)_{T}=\left(\frac{\partial^2 P}{\partial R^2_A}\right)_{T}=0\,,
\end{equation}
and one can easily check that this kind of solution does not exist. Furthermore,
from the first equation in (\ref{PVTc}), we get
\begin{eqnarray}\label{dPVTB}
\left.\frac{\partial P}{\partial R_{A}}\right|_{T}=-\frac{1+2\pi R_{A}T}{4\pi R_{A}^3}<0 \,,
\end{eqnarray}
which implies that the system is thermodynamically stable.

The stability of the thermodynamic system can also be understood by investigating the specific heat capacity at constant pressure $C_{P}$, which is positive for a stable system and negative for a unstable system. By using the entropy in (\ref{MSVWdef}) and  (\ref{PVT}), we obtain the specific heat capacity of the FRW universe
\begin{eqnarray}\label{CP}
C_{P}=\left(\frac{\partial {\cal H}}{\partial T}\right)_{P}=T\left(\frac{\partial S}{\partial T}\right)_{P}=\frac{2{\pi}R_{A}^{2}(-1+8{\pi}R_{A}^2 P)}{1+8{\pi} R_{A}^2 P}=\frac{4{\pi}^{2}R_{A}^{3} T}{1+2{\pi}T R_{A}}\,,
\end{eqnarray}
where ${\cal H}=U+P V$ is the enthalpy of the system.
Clearly one sees that the sign of the heat capacity $C_P$ is always positive since the temperature should take positive
value for $1/\sqrt{8{\pi} P}<R_{A}$, implying the FRW universe is thermally stable,
while many black holes have negative heat capacities and thus unstable. The stable FRW universe also provides a suitable background for the black hole formation,
such as the McVittie black hole \cite{Abdusattar:2022bpg}. The expression of the heat capacity (\ref{CP}) can be also used to conveniently calculate the Joule-Thomson coefficient as shown in the following section.

\section{Joule-Thomson expansion for a FRW universe}\label{sIV}

In this section, by using the above thermodynamic equation of state for a FRW universe with the cosmological constant, we investigate its Joule-Thomson expansion, and obtain the Joule-Thomson coefficient. Furthermore, we will demonstrate whether the inversion temperature and inversion pressure exist.

The Joule-Thomson expansion is an interesting physical process with important feature that the temperature changes with pressure while keeping the enthalpy ${\cal H}$ fixed during this expansion process \cite{Okcu:2016tgt}.
The Joule-Thomson coefficient $\mu$ is defined by
\begin{equation}\label{JTC0}
\mu=\large \left(\frac{\partial T}{\partial P}\large \right)_{\cal H} \, .
\end{equation}
The sign of $\mu$ is determined by cooling or heating occurred in the thermodynamic system. $\mu>0$ means that the system remains cooling, and the temperature of the system decreases. On the contrary, $\mu<0$ means that the system remains heating. Therefore, the cooling-heating inversion points lie at $\mu=0$, and the temperature of the thermodynamic system at that point is named as inversion temperature $T_i$. When the temperature of the system is $T_i$, the corresponding pressure is considered as the inversion pressure $P_i$, and consequently the special point ($T_i$, $P_i$) is called the inversion point. The (\ref{JTC0}) is rewritten as following form \cite{Okcu:2016tgt}
\begin{equation}\label{mu1}
\mu=\frac{1}{{C}_{P}}\left[T\left(\frac{\partial V}{\partial T}\right)_{P}-V \right]\,.
\end{equation}

The enthalpy of FRW universe is given by
\begin{equation}\label{HH}
 {\cal H}=\frac{R_{A}(2\pi R_{A} T-1)}{3}=-\frac{R_A}{2}+\frac{4\pi R_A^3}{3}P \,.
\end{equation}
One can see that the enthalpy ${\cal H}>0$ when
\begin{equation}\label{HH1}
R_A>\frac{1}{2\pi T}~~~~~ or ~~~~~ R_A>\sqrt{\frac{3}{8\pi P}}\,,
\end{equation}
and the enthalpy ${\cal H}<0$ when
\begin{equation}\label{HH2}
 0<R_A<\frac{1}{2\pi T}~~~~~ or ~~~~~ \frac{1}{\sqrt{8\pi P}}<R_A<\sqrt{\frac{3}{8\pi P}}\,.
\end{equation}

From (\ref{HH}), the pressure $P$ can be rewritten as a function of ${\cal H}$ and $R_A$,
\begin{equation} \label{PHRA}
P({\cal H},R_A)=\frac{3(2{\cal H}+R_A)}{8\pi R_A^3} \,,
\end{equation}
and then substituting the Eq.(\ref{PHRA}) into (\ref{PVT}), one writes the temperature as a function of ${\cal H}$ and $R_A$ as follows
\begin{equation} \label{THRA}
T({\cal H},R_A)=\frac{3{\cal H}+R_A}{2\pi R_A^2}\,.
\end{equation}
By using (\ref{JTC0}) combined with Eqs.(\ref{PHRA}) and (\ref{THRA}), we obtain the Joule-Thomson coefficient of the FRW universe as a function of $T$ and $R_A$,
\begin{equation}\label{JTE1}
\mu=\large \frac{\left({\partial T}/{\partial R_A}\large \right)_{{\cal H}}}{\large \left({\partial P}/{\partial R_A}\large \right)_{{\cal H}}}=\frac{2R_A(R_A+6 \cal H)}{3(R_A+3 \cal H)}=\frac{4 R_{A}}{3}-\frac{1}{3\pi T} \,,
\end{equation}
which can also be obtain by using Eqs.(\ref{PVT}), (\ref{CP}) and (\ref{mu1}).
We can see from the above equation, if the value of enthalpy ${\cal H}$ is positive (${\cal H}>0$), suggesting that the FRW universe is always in the cooling stage, i.e., there is no inversion temperature and inversion pressure. If ${\cal H}$ is negative (${\cal H}<0$), there exist both a divergence point and an inversion point of FRW universe.\footnote{Note that, in the same way as the internal energy $U$ being dependent on the choice of $U_0$ (explained in Footnote \ref{ftn3}), the enthalpy ${\cal H}$ also depends on $U_0$, i.e.\ we can alternatively define the enthalpy as $\widetilde {\cal H}=U+U_0+PV={\cal H}+U_0$. However, we can easily prove that the JT expansion coefficient is independent of the choice of $U_0$, i.e. $\mu =\frac{2R_A[R_A+6 (\widetilde{\cal H}-U_0)]}{3[R_A+3 (\widetilde{\cal H}-U_0)]}$ same as Eq.(\ref{JTE1}). Furthermore, the condition for the inversion point to exist is also the same, i.e.\ the condition is $\widetilde{\cal H}-U_0={\cal H}<0$. As an additional remark, we can further easily derive that the condition ${\cal H}<0$ is equivalent to $\rho+p>0$, which is just the null energy condition and can be satisfied by most matter fields.} The inversion point
corresponds to $\ddot{a}=0$, which derivations and physical interpretations are written in the Appendix \ref{appA}.
Setting $\mu=0$, and combining with (\ref{HH}), we get the inversion temperature is given by
\begin{eqnarray}\label{invT}
T_{i}=\frac{1}{4\pi R_A}=-\frac{1}{24\pi \cal H}\,.
\end{eqnarray}
Combining with (\ref{PVT}) and (\ref{HH}), we get the inversion pressure
\begin{equation}\label{invP}
P_{i}=\frac{1}{4\pi R_A^2}=\frac{1}{144\pi {\cal H}^2}\,.
\end{equation}
From the above two relations, we easily find that the inversion temperature and inversion pressure both exist if the enthalpy ${\cal H}$ for a FRW universe is negative. In addition, from (\ref{invT}) and (\ref{invP}), we also obtain the simple relation between the inversion temperature and the inversion pressure
\begin{eqnarray}\label{invTP}
T_{i}=\frac{1}{2}\sqrt{\frac{P_{i}}{\pi}}\,.
\end{eqnarray}

\begin{figure}[h]
\centering
 \begin{minipage}[t]{1\linewidth}
\includegraphics[width=8cm]{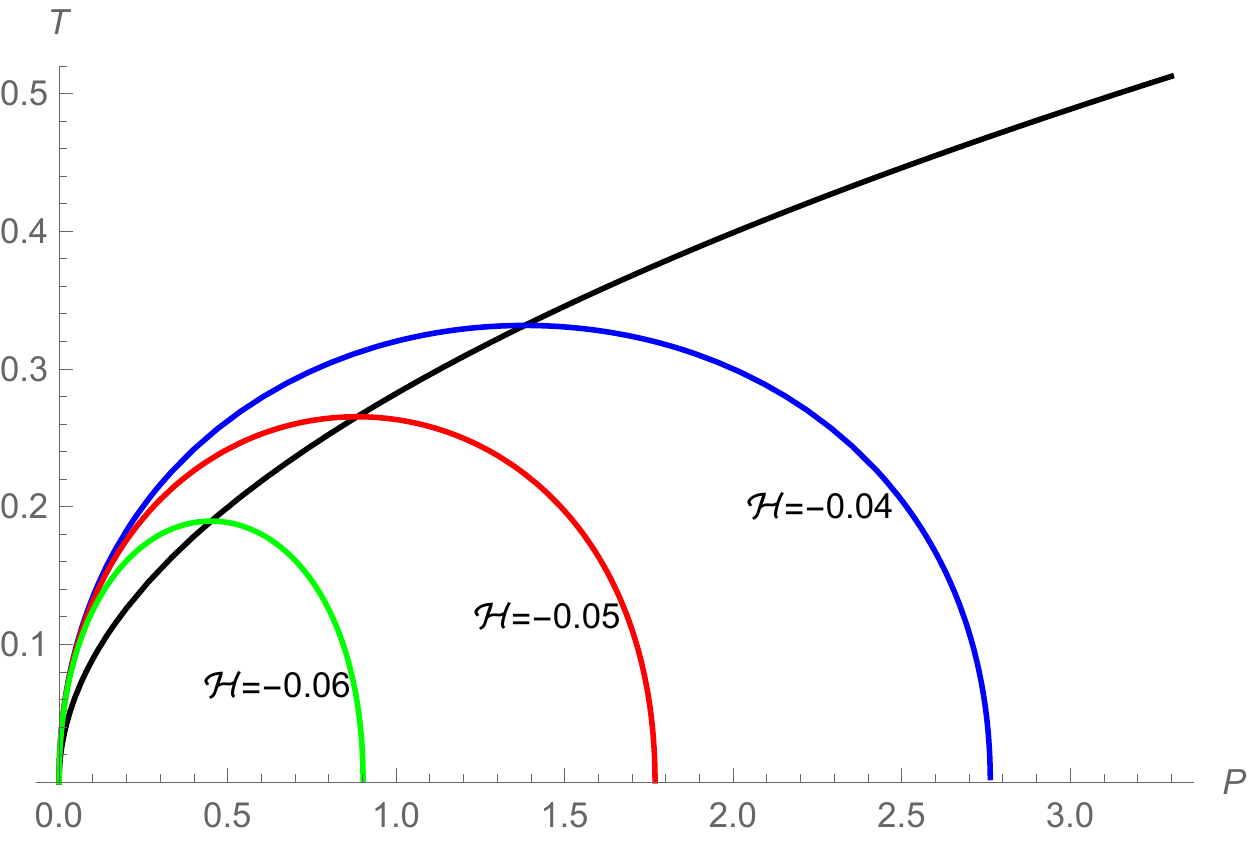}
\end{minipage}
 \caption{\label{FigJT} \footnotesize The isenthalpic curves for a FRW universe have been plotted with different values of ${\cal H}$, while the monotonous black line is the inversion curve. These isenthalpic curves are separated by the inversion curve into two regions, i.e. the left cooling and right heating regions.}
\end{figure}

The Joule-Thomson expansion occurs as an isenthalpic process, and hence it is significant to study the isenthalpic curves for a thermodynamic system. From Eqs.(\ref{PHRA}), (\ref{THRA}) and (\ref{invTP}), we will plot the isenthalpic curves in the Fig.\ref{FigJT} for a FRW universe in the $T$-$P$ plane by fixing the ${\cal H}$, and also show the inversion curve.\footnote{In Fig.\ref{FigJT}, we have plotted isenthalpic curves to exhibit the cooling-heating property, and an inversion curve intersects at their maxima. Note that fixing the enthalpy ${\cal H}$ is a necessary condition for cooling-heating, but it does not mean that the evolution of the universe is isenthalpic. In general, as the universe is dynamical, its enthalpy ${\cal H}$, pressure $P$ and temperature $T$ change with time. As long as they satisfy the relation indicated in (\ref{invTP}), transition from cooling to heating or from heating to cooling can happen.}

Obviously, these isenthalpic curves intersect with the inversion curve at their maximum points. In addition, these isenthalpic curves are also separated into two regions by the inversion curve, i.e. the left cooling and right heating regions. Furthermore, the region surrounded by isenthalpic curve and $P$ axis shrinks when ${\cal H}$ decreases or becomes more negative.

\section{Conclusions and Discussion}\label{sV}

In view of limitations of application of event horizon in realistic astrophysics, black hole thermodynamics associated to apparent horizon is developed. This point is especially significant for the Universe, since under several situations the event horizon does not exist. The unified first law of thermodynamics concentrates on apparent horizon, and thus can be naturally applied to cosmology.

By using the unified first law we set up the equation of state of the FRW universe. The thermodynamic pressure $P$ is determined by the unified first law, in fact the field equation, directly.  Different from the AdS black holes, we find that the proper thermodynamic pressure $P$ for a FRW universe involves the work density $W$ of the matter field and the cosmological constant $\Lambda$. Theoretically, this definition of $P$ is the conjugate quantity of thermodynamic volume. As a comparison, the widely used definition thermodynamic pressure in investigations in extended space of black holes is only an analogy in dynamics of cosmology. A rigorous derivation from fundamental laws is deficient. Furthermore, we find the specific capacity of the FRW universe is positive. This result is remarkable, since it implies that the FRW universe is stable. Contrarily, the instability of black hole against thermodynamics perturbations is an eminent progress in black hole thermodynamics,  since the specific capacity of a black hole is negative.

After careful calculations we formulate the thermodynamic pressure as a function of apparent horizon and temperature  $P=P(R_A, T)$, and apply it to the studies of Joule-Thomson expansion for a FRW universe. We demonstrate that the FRW universe has an inversion temperature if its enthalpy is negative, and clearly show the cooling and heating phase in a plot.

Despite the findings in this paper, there are still many open questions. A natural question is whether our definition of the thermodynamic pressure can be used in other spacetime besides the FRW universe, e.g.\ black holes immersed in perfect fluids. From the viewpoint of the new definition of $P$, there may be some underlying physical interpretations, how to further understand the profound physical implication between $P$ and $P_{\Lambda}$ will be an intriguing work. Another interesting question is whether the cooling-heating property can be found for the FRW universe in modified theories of gravity and/or filled with other matter fields instead of perfect fluid, and the results may provide new points to test the modified theories of gravity.
We are also very curious about whether the inversion temperature can be found by cosmological observations, which will be a challenging and meaningful task. These questions are studied in the future.
Interestingly, as discussed in Appendix \ref{appA}, we have investigated the constraint on the perfect fluid from $T\geq0$ and the physical meaning of the inversion point.
We have obtained that at the inversion point $\mu=0$ is equivalent to $\ddot{a}=0$, corresponding to the acceleration/deceleration transition of the universe, which to our knowledge is discovered for the first time in standard cosmology.

\section{Acknowledgment}

We would like to thank the anonymous referee for helpful comments and Profs. Li-Ming Cao, Shao-Wen Wei, Xiao-Mei Kuang, Yen Chin Ong, Yihao Yin and  Theodore A. Jacobson for useful discussions. This work is supported by National Natural Science Foundation of China (NSFC) under grant Nos. 12175105, 12147175, 12235019,12174194, 11575083, Top-notch Academic Programs Project of Jiangsu Higher Education Institutions (TAPP). W.-L. Y acknowledges the startup fund of Nanjing University of
Aeronautics and Astronautics under grant No. 1008-YAH20006 and stable support for basic institute research (Grant No. 190101). H.Z. is supported by Shandong Province Natural Science Foundation under grant No. ZR201709220395, and the National Key Research and Development Program of China (No. 2020YFC2201400).

\appendix

\section{Constraints from $\kappa<0$ and $\mu=0$}\label{appA}

In this appendix, we first give the constraint on the perfect fluid from $\kappa<0$, and then give
the constraint on the scale factor from the inversion point ($\mu=0$).

\bigskip

From the expression (\ref{sg}) of the surface gravity $\kappa$, one can see that $\kappa<0$ is equivalent to
\begin{equation}
1-\frac{\dot{R}_A}{2HR_A}>0\,,
\end{equation}
which together with (\ref{p}) results to
\begin{equation}
p-\frac{\Lambda}{8\pi}-\frac{1}{8\pi R_A^2}<0\,,
\end{equation}
and by using (\ref{rho}), we finally get
\begin{equation}
\rho-3p>-\frac{\Lambda}{2\pi}\,.  \label{A2}
\end{equation}

For the cosmological horizon of the McVittie spacetime, $\kappa<0$ is equivalent to \cite{Abdusattar:2022bpg}
\begin{equation}
\rho-3p>\frac{m}{2\pi R_{Ac}^2}\,, \label{A3}
\end{equation}
where $m$ is the mass parameter of the central black hole, $R_{Ac}$ is the larger apparent horizon (cosmological horizon)
of the McVittie spacetime, and the cosmological constant is set to zero in that paper.
If one sets the mass parameter of the McVittie spacetime to be zero and recovers the cosmological constant,
one will get a spatially flat FRW universe with a cosmological constant, and condition (\ref{A3}) can reduce
to condition (\ref{A2}), i.e. the conditions for $\kappa<0$ in the two papers are consistent.

\bigskip

For the $\mathcal{H}<0$ case, the existence of an inversion point (a maximum in the $T$-$P$ diagram) in the JT expansion is very interesting, and its physical meaning needs to be further excavated. We find that at this point,
the acceleration of the FRW universe is zero, i.e. $\ddot{a}=0$, which derivations are simple and shown below.

Equate the Hawking temperature (\ref{T1}) and the inversion temperature (\ref{invT}), one can get
\begin{equation}
\dot{R}_A-H R_A=0\,,
\end{equation}
which together with (\ref{dAH}) results to the following relation
\begin{equation}
R_A^2\left(\dot{H}-\frac{k}{a^2}\right)+1=0\,,
\end{equation}
and by using (\ref{AH}), we finally get
\begin{equation}
\dot{H}+H^2=0\,, \label{B2}
\end{equation}
i.e.
\begin{equation}
\ddot{a}=0\,. \label{B3}
\end{equation}

This result shows that the second-derivative of the scale factor of the FRW universe is zero at the inversion point,
which means that its signs are different for $T<T_i$ and $T>T_i$, i.e. there is a transition from acceleration $\ddot{a}>0$
to deceleration $\ddot{a}<0$ or from deceleration $\ddot{a}<0$ to acceleration $\ddot{a}>0$.

The above condition (\ref{B2}) or (\ref{B3}) also corresponds to a constraint on the perfect fluid, which is still obtained by using equations (\ref{AH}), (\ref{dAH}), (\ref{rho}) and (\ref{p}), and the constraint is
\begin{equation}
\rho+3p=\frac{\Lambda}{4\pi}\,.
\end{equation}
When the temperature is larger than the inversion temperature, the perfect fluid satisfies $\rho+3p<\Lambda/(4\pi)$,
and when the temperature is smaller than the inversion temperature, the perfect fluid satisfies $\rho+3p>\Lambda/(4\pi)$,
so one can also say that there is a phase transition of different fluids at the inversion point.

The inversion point signals a new type of phase transition that is discovered for the first time to our knowledge, and we are very interested whether it can be matched with observations of our real universe.


\begin{thebibliography}{11}

\bibitem{Hawking:1974sw}
S.~W.~Hawking,
``Particle Creation by Black Holes",
{\hypersetup{urlcolor=vividviolet}\href{https://inspirehep.net/literature/101338}{Commun. Math. Phys. \textbf{43}, 199-220 (1975)}}, \href{https://inspirehep.net/literature/101338}{[erratum: Commun. Math. Phys. \textbf{46}, 206 (1976)]}.

\bibitem{Bardeen1973}
J. M. Bardeen, B. Carter and S. W. Hawking,
``The Four laws of black hole mechanics",
{\hypersetup{urlcolor=vividviolet}\href{https://inspirehep.net/literature/81181}{Commun. Math. Phys. 31, 161 (1973)}}.

\bibitem{Bekenstein:1974ax}
J.~D.~Bekenstein,``Generalized second law of thermodynamics in black hole physics",
{\hypersetup{urlcolor=vividviolet}\href{https://inspirehep.net/literature/91796}{Phys. Rev. D \textbf{9}, 3292-3300 (1974)}}.

\bibitem{Cai:2005ra}
R.~G.~Cai and S.~P.~Kim,
``First law of thermodynamics and Friedmann equations of Friedmann-Robertson-Walker universe'',
{\hypersetup{urlcolor=vividviolet}\href{https://inspirehep.net/literature/674617}{JHEP \textbf{02}, 050 (2005)}},
\href{https://arxiv.org/abs/hep-th/0501055v1}{[arXiv:hep-th/0501055]}.

\bibitem{Gong:2007md}
Y.~Gong and A.~Wang,
``The Friedmann equations and thermodynamics of apparent horizons'',
{\hypersetup{urlcolor=vividviolet}\href{https://inspirehep.net/literature/748131}{Phys. Rev. Lett. \textbf{99}, 211301 (2007)}}, \href{https://arxiv.org/abs/0704.0793}{[arXiv:hep-th/0704.0793]}.

\bibitem{Gibbons:1977mu}
G.~W.~Gibbons and S.~W.~Hawking,
``Cosmological Event Horizons, Thermodynamics, and Particle Creation",
{\hypersetup{urlcolor=vividviolet}\href{https://journals.aps.org/prd/abstract/10.1103/PhysRevD.15.2738}{Phys. Rev. D \textbf{15}, 2738-2751 (1977)}}.


\bibitem{Hayward:1997jp}
S.~A.~Hayward,
``Unified first law of black hole dynamics and relativistic thermodynamics'',
{\hypersetup{urlcolor=vividviolet}\href{https://inspirehep.net/literature/449938}{Class. Quant. Grav. \textbf{15}, 3147-3162 (1998)}},
\href{https://arxiv.org/abs/gr-qc/9710089}{[arXiv:gr-qc/9710089]};
S.~A.~Hayward,
{\hypersetup{urlcolor=vividviolet}\href{https://journals.aps.org/prd/abstract/10.1103/PhysRevD.49.6467}{Phys. Rev. D \textbf{49}, 6467-6474 (1994)}}.

\bibitem{Cai:2006rs}
R.~G.~Cai and L.~M.~Cao,
``Unified first law and thermodynamics of apparent horizon in FRW universe'',
{\hypersetup{urlcolor=vividviolet}\href{https://inspirehep.net/literature/731742}{Phys. Rev. D \textbf{75}, 064008
(2007)}}, \href{https://arxiv.org/pdf/gr-qc/0611071}{[arXiv:gr-qc/0611071]}.

\bibitem{Maeda:2007uu}
H.~Maeda and M.~Nozawa,%
``Generalized Misner-Sharp quasi-local mass in Einstein-Gauss-Bonnet gravity'',
{\hypersetup{urlcolor=vividviolet}\href{https://inspirehep.net/literature/760308}{Phys. Rev. D \textbf{77} (2008), 064031}},
\href{https://arxiv.org/abs/0709.1199v4}{[arXiv:hep-th/0709.1199]}.

\bibitem{Cai:2009qf}
R.~G.~Cai, L.~M.~Cao, Y.~P.~Hu, and N.~Ohta,
``Generalized Misner-Sharp Energy in f(R) Gravity'',
{\hypersetup{urlcolor=vividviolet}\href{https://inspirehep.net/literature/833807}{Phys. Rev. D \textbf{80} (2009), 104016}},
\href{https://arxiv.org/abs/0910.2387}{[arXiv:hep-th/0910.2387]}.




\bibitem{Hu:2015xva}
Y.~P.~Hu and H.~Zhang,
``Misner-Sharp Mass and the Unified First Law in Massive Gravity'',
{\hypersetup{urlcolor=vividviolet}\href{https://inspirehep.net/literature/1342429}{Phys. Rev. D {\bf 92}, no.2, 024006 (2015)}},
\href{https://arxiv.org/abs/1502.00069}{[arXiv:hep-th/1502.00069]}.

\bibitem{Kubiznak:2012wp}
D.~Kubiznak and R.~B.~Mann,
``$P$-$V$ criticality of charged AdS black holes'',
{\hypersetup{urlcolor=vividviolet}\href{https://inspirehep.net/literature/1113435}{JHEP \textbf{07}, 033 (2012)}},
\href{https://arxiv.org/pdf/1205.0559}{[arXiv:hep-th/1205.0559]}.


\bibitem{Xu:2015rfa}
J.~Xu, L.~M.~Cao and Y.~P.~Hu,
``$P$-$V$ criticality in the extended phase space of black holes in massive gravity'',
{\hypersetup{urlcolor=vividviolet}\href{https://inspirehep.net/literature/1375807}{Phys. Rev. D \textbf{91}, no.12, 124033 (2015)}},
\href{https://arxiv.org/pdf/1506.03578}{[arXiv:gr-qc/1506.03578]};
R.~G.~Cai, Y.~P.~Hu, Q.~Y.~Pan and Y.~L.~Zhang,
``Thermodynamics of Black Holes in Massive Gravity'',
{\hypersetup{urlcolor=vividviolet}\href{https://inspirehep.net/literature/1315428}{Phys. Rev. D \textbf{91}, no.2, 024032 (2015)
}}, \href{https://arxiv.org/abs/1409.2369}{[arXiv:hep-th/1409.2369]}.


\bibitem{Kubiznak:2016qmn}
D.~Kubiznak, R.~B.~Mann and M.~Teo,
``Black hole chemistry: thermodynamics with Lambda'',
{\hypersetup{urlcolor=vividviolet}\href{https://inspirehep.net/literature/1482749}{Class. Quant. Grav. \textbf{34}, no.6, 063001
(2017)}}, \href{https://arxiv.org/abs/1608.06147}{[arXiv:hep-th/1608.06147]};
D.~Kubiznak and R.~B.~Mann,
{\hypersetup{urlcolor=vividviolet}\href{https://inspirehep.net/literature/1289053}{Can. J. Phys. \textbf{93}, no.9, 999-1002 (2015)}}, \href{https://arxiv.org/abs/1404.2126}{[arXiv:gr-qc/1404.2126]}.

\bibitem{Hu:2018qsy}
Y.~P.~Hu, H.~A.~Zeng, Z.~M.~Jiang and H.~Zhang,
``$P$-$V$ criticality in the extended phase space of black holes in Einstein-Horndeski gravity'',
{\hypersetup{urlcolor=vividviolet}\href{https://inspirehep.net/literature/1711378}{Phys. Rev. D \textbf{100}, no.8, 084004
(2019)}}, \href{https://arxiv.org/pdf/1812.09938}{[arXiv:gr-qc/1812.09938]}.

\bibitem{Gunasekaran:2012dq}
S.~Gunasekaran, R.~B.~Mann and D.~Kubiznak,
``Extended phase space thermodynamics for charged and rotating black holes and Born-Infeld vacuum polarization",
{\hypersetup{urlcolor=vividviolet}\href{https://inspirehep.net/literature/1183854}{JHEP \textbf{11}, 110 (2012)}}, \href{https://arxiv.org/abs/1208.6251}{[arXiv:hep-th/1208.6251]}.

\bibitem{Wei:2012ui}
S.~W.~Wei and Y.~X.~Liu,
``Critical phenomena and thermodynamic geometry of charged Gauss-Bonnet AdS black holes'',
{\hypersetup{urlcolor=vividviolet}\href{https://inspirehep.net/literature/1184929}{Phys. Rev. D \textbf{87}, no.4, 044014 (2013)}},
\href{https://arxiv.org/abs/1209.1707}{[arXiv:gr-qc/1209.1707]};
P.~Cheng, S.~W.~Wei and Y.~X.~Liu,
{\hypersetup{urlcolor=vividviolet}\href{https://inspirehep.net/literature/1436347}{Phys. Rev. D \textbf{94} (2016), 024025}},
\href{https://arxiv.org/abs/1603.08694}{[arXiv:gr-qc/1603.08694]};
S.~W.~Wei and Y.~X.~Liu,
{\hypersetup{urlcolor=vividviolet}\href{https://inspirehep.net/literature/1789085}{Phys. Rev. D \textbf{101}, no.10, 104018 (2020)}}, \href{https://arxiv.org/abs/2003.14275}{[arXiv:gr-qc/2003.14275]}.

\bibitem{Hendi:2012um}
S.~H.~Hendi and M.~H.~Vahidinia,
``Extended phase space thermodynamics and $P$-$V$ criticality of black holes with a nonlinear source'',
{\hypersetup{urlcolor=vividviolet}\href{https://inspirehep.net/literature/1208685}{Phys. Rev. D \textbf{88}, no.8, 084045 (2013)}}, \href{https://arxiv.org/abs/1212.6128}{[arXiv:hep-th/1212.6128]};
S.~H.~Hendi, R.~B.~Mann, S.~Panahiyan and B.~Eslam Panah,
{\hypersetup{urlcolor=vividviolet}\href{https://inspirehep.net/literature/1711378}{Phys. Rev. D \textbf{95}, no.2, 021501 (2017)}}, \href{https://arxiv.org/pdf/1812.09938}{[arXiv:gr-qc/1702.00432]}.

\bibitem{Cai:2013qga}
R.~G.~Cai, L.~M.~Cao, L.~Li and R.~Q.~Yang,
``$P$-$V$ criticality in the extended phase space of Gauss-Bonnet black holes in AdS space",
{\hypersetup{urlcolor=vividviolet}\href{https://inspirehep.net/literature/1239956}{JHEP \textbf{09}, 005 (2013)}}, \href{https://arxiv.org/abs/1306.6233}{[arXiv:gr-qc/1306.6233]}.


\bibitem{Altamirano:2013ane}
N.~Altamirano, D.~Kubiznak and R.~B.~Mann,
``Reentrant phase transitions in rotating anti\textendash{}de Sitter black holes'',
{\hypersetup{urlcolor=vividviolet}\href{https://inspirehep.net/literature/1239816}{Phys. Rev. D \textbf{88}, no.10, 101502 (2013)
}}, \href{https://arxiv.org/abs/1306.5756}{[arXiv:hep-th/1306.5756]};
N.~Altamirano, D.~Kubiznak, R.~B.~Mann and Z.~Sherkatghanad,
{\hypersetup{urlcolor=vividviolet}\href{https://inspirehep.net/literature/1276853}{Galaxies \textbf{2} (2014), 89-159}}, \href{https://arxiv.org/abs/1401.2586}{[arXiv:hep-th/1401.2586]}.

\bibitem{Bhattacharya:2017nru}
K.~Bhattacharya, B.~R.~Majhi and S.~Samanta,
``Van der Waals criticality in AdS black holes: a phenomenological study'',
{\hypersetup{urlcolor=vividviolet}\href{https://inspirehep.net/literature/1622513}{Phys. Rev. D \textbf{96}, no.8, 084037 (2017)}}, \href{https://arxiv.org/abs/1709.02650}{[arXiv:gr-qc/1709.02650]}.

\bibitem{Li:2020xkh}
R.~Li and J.~Wang,
``Hawking radiation and $P$-$V$ criticality of charged dynamical (Vaidya) black hole in anti-de Sitter space", {\hypersetup{urlcolor=vividviolet}\href{https://inspirehep.net/literature/1818176}{Phys. Lett. B \textbf{813}, 136035 (2021)}}, \href{https://arxiv.org/abs/2009.09319}{[arXiv:gr-qc/2009.09319]}.

\bibitem{Hu:2020pmr}
Y.~P.~Hu, L.~Cai, X.~Liang, S.~B.~Kong and H.~Zhang,
``Divergence behavior of thermodynamic curvature scalar at critical point in the extended phase space of generic black holes'',
{\hypersetup{urlcolor=vividviolet}\href{https://inspirehep.net/literature/1823798}{Phys. Lett. B \textbf{822}, 136661 (2021)}}, \href{https://arxiv.org/abs/2010.09363}{[arXiv:gr-qc/2010.09363]}.


\bibitem{Kastor:2009wy}
D.~Kastor, S.~Ray, and J.~Traschen,
``Enthalpy and the Mechanics of AdS Black Holes'',
{\hypersetup{urlcolor=vividviolet}\href{https://inspirehep.net/literature/818236}{Class. Quant. Grav. \textbf{26}, 195011 (2009)}},
\href{https://arxiv.org/pdf/0904.2765}{[arXiv:hep-th/0904.2765]}.

\bibitem{Dolan:2010ha}
B.~P.~Dolan,``The cosmological constant and the black hole equation of state",
{\hypersetup{urlcolor=vividviolet}\href{https://inspirehep.net/literature/866593}{Class. Quant. Grav. \textbf{28}, 125020 (2011)}}, \href{https://arxiv.org/pdf/1008.5023}{[arXiv:gr-qc/1008.5023]};
B.~P.~Dolan,
{\hypersetup{urlcolor=vividviolet}\href{https://inspirehep.net/literature/916548}{Class. Quant. Grav. \textbf{28}, 235017 (2011)}}, \href{https://arxiv.org/abs/1106.6260}{[arXiv:gr-qc/1106.6260]}.

\bibitem{Cvetic:2010jb}
M.~Cvetic, G.~W.~Gibbons, D.~Kubiznak and C.~N.~Pope,
``Black Hole Enthalpy and an Entropy Inequality for the Thermodynamic Volume'',
{\hypersetup{urlcolor=vividviolet}\href{https://inspirehep.net/literature/881245}{Phys. Rev. D \textbf{84}, 024037 (2011)}}, \href{https://arxiv.org/abs/1012.2888}{[arXiv:hep-th/1012.2888]}.

\bibitem{Debnath:2020inx}
U.~Debnath,``Thermodynamics of FRW Universe: Heat Engine'',
{\hypersetup{urlcolor=vividviolet}\href{https://inspirehep.net/literature/1821427}{Phys. Lett. B \textbf{810}, 135807 (2020)}},
\href{https://arxiv.org/pdf/2010.02102}{[arXiv:gr-qc/2010.02102]}.


\bibitem{Jacobson:1995ab}
T.~Jacobson,``Thermodynamics of space-time: The Einstein equation of state'',
{\hypersetup{urlcolor=vividviolet}\href{https://inspirehep.net/literature/394001}{Phys. Rev. Lett. \textbf{75}, 1260-1263 (1995)}},
\href{https://arxiv.org/pdf/gr-qc/9504004}{[arXiv:gr-qc/9504004]}.

\bibitem{Cai:2008ys}
R.~G.~Cai, L.~M.~Cao and Y.~P.~Hu,
``Corrected Entropy-Area Relation and Modified Friedmann Equations'',
{\hypersetup{urlcolor=vividviolet}\href{https://inspirehep.net/literature/394001}{JHEP \textbf{08}, 090 (2008)}},
\href{https://arxiv.org/abs/0807.1232}{[arXiv:hep-th/0807.1232]}.

\bibitem{Akbar:2006kj}
M.~Akbar and R.~G.~Cai,
``Thermodynamic Behavior of Friedmann Equations at Apparent Horizon of FRW Universe'',
{\hypersetup{urlcolor=vividviolet}\href{https://inspirehep.net/literature/726509}{Phys. Rev. D \textbf{75}, 084003 (2007)}},
\href{https://arxiv.org/pdf/hep-th/0609128}{[arXiv:hep-th/0609128]}.

\bibitem{Akbar:2006mq}
M.~Akbar and R.~G.~Cai,
``Thermodynamic Behavior of Field Equations for f(R) Gravity'',
{\hypersetup{urlcolor=vividviolet}\href{https://inspirehep.net/literature/734489}{Phys. Lett. B \textbf{648}, 243-248 (2007)}},
\href{https://arxiv.org/abs/gr-qc/0612089}{[arXiv:gr-qc/0612089]}.

\bibitem{Zhang:2014ala}
H.~Zhang and X.~Z.~Li,
``From thermodynamics to the solutions in gravity theory",
{\hypersetup{urlcolor=vividviolet}\href{https://inspirehep.net/literature/1299558}{Phys. Lett. B \textbf{737}, 395-400 (2014)}},
\href{https://arxiv.org/abs/1406.1553}{[arXiv:gr-qc/1406.1553]};
H.~Zhang, S.~A.~Hayward, X.~H.~Zhai and X.~Z.~Li,
{\hypersetup{urlcolor=vividviolet}\href{https://inspirehep.net/literature/1294611}{Phys. Rev. D \textbf{89}, no.6, 064052 (2014)}},
\href{https://arxiv.org/abs/1304.3647}{[arXiv:gr-qc/1304.3647]}.

\bibitem{Kong:2021qiu}
S.~B.~Kong, H.~Abdusattar, Y.~Yin and Y.~P.~Hu,
``The $P$-$V$ phase transition of the FRW universe'',
{\hypersetup{urlcolor=vividviolet}\href{https://inspirehep.net/literature/2512592}{Eur. Phys. J. C \textbf{82} (2022) no.11, 1047}},
\href{https://arxiv.org/abs/2108.09411}{[arXiv:gr-qc/2108.09411]}.

\bibitem{Okcu:2016tgt}
\"O.~\"Okc\"u and E.~Ayd\i{}ner,
``Joule\textendash{}Thomson expansion of the charged AdS black holes'',
{\hypersetup{urlcolor=vividviolet}\href{https://inspirehep.net/literature/1499457}{Eur. Phys. J. C \textbf{77}, no.1, 24 (2017)}},
\href{https://arxiv.org/pdf/1611.06327}{[arXiv:gr-qc/1611.06327]}.

\bibitem{Okcu:2017qgo}
\"O.~\"Okc\"u and E.~Ayd\i{}ner,
``Joule\textendash{}Thomson expansion of Kerr\textendash{}AdS black holes'',
{\hypersetup{urlcolor=vividviolet}\href{https://inspirehep.net/literature/1624404}{Eur. Phys. J. C \textbf{78}, no.2, 123 (2018)}},
\href{https://arxiv.org/abs/1709.06426}{[arXiv:gr-qc/1709.06426]}.

\bibitem{Lan:2018nnp}
S.~Q.~Lan,
``Joule-Thomson expansion of charged Gauss-Bonnet black holes in AdS space'',
{\hypersetup{urlcolor=vividviolet}\href{https://inspirehep.net/literature/1673217}{Phys. Rev. D \textbf{98}, no.8, 084014 (2018)}},
\href{https://arxiv.org/pdf/1805.05817.pdf}{[arXiv:gr-qc/1805.05817]}.

\bibitem{Pu:2019bxf}
J.~Pu, S.~Guo, Q.~Q.~Jiang, and X.~T.~Zu,
``Joule-Thomson expansion of the regular(Bardeen)-AdS black hole'',
{\hypersetup{urlcolor=vividviolet}\href{https://inspirehep.net/literature/1733730}{Chin. Phys. C \textbf{44}, no.3, 035102 (2020)}},
\href{https://arxiv.org/pdf/1905.02318.pdf}{[arXiv:gr-qc/1905.02318]}.

\bibitem{Li:2019jcd}
C.~Li, P.~He, P.~Li and J.~B.~Deng,
``Joule-Thomson expansion of the Bardeen-AdS black holes'',
{\hypersetup{urlcolor=vividviolet}\href{https://inspirehep.net/literature/1730522}{Gen. Rel. Grav. \textbf{52}, no.5, 50
(2020)}}, \href{https://arxiv.org/abs/1904.09548}{[arXiv:gr-qc/1904.09548]}.

\bibitem{Rajani:2020mdw}
K.~V.~Rajani, C.~L.~A.~Rizwan, A.~Naveena Kumara, M.~S.~Ali and D.~Vaid,
``Joule\textendash{}Thomson expansion of regular Bardeen AdS black hole surrounded by static anisotropic matter field'',
{\hypersetup{urlcolor=vividviolet}\href{https://inspirehep.net/literature/1779480}{Phys. Dark Univ. \textbf{32}, 100825
(2021)}}, \href{https://arxiv.org/abs/2002.03634}{[arXiv:gr-qc/2002.03634]}.

\bibitem{Bi:2020vcg}
S.~Bi, M.~Du, J.~Tao, and F.~Yao,
``Joule-Thomson expansion of Born-Infeld AdS black holes'',
{\hypersetup{urlcolor=vividviolet}\href{https://inspirehep.net/literature/1801469}{Chin. Phys. C \textbf{45}, no.2, 025109 (2021)}},
\href{https://arxiv.org/pdf/2006.08920.pdf}{[arXiv:gr-qc/2006.08920]}.

\bibitem{Ghaffarnejad:2018exz}
H.~Ghaffarnejad, E.~Yaraie and M.~Farsam,
``Quintessence Reissner Nordstr\"om Anti de Sitter Black Holes and Joule Thomson effect'',
{\hypersetup{urlcolor=vividviolet}\href{https://inspirehep.net/literature/1657412}{Int. J. Theor. Phys. \textbf{57} (2018) no.6, 1671-1682}},
\href{https://arxiv.org/abs/1802.08749}{[arXiv:gr-qc/1802.08749]};
M.~Chabab, H.~El Moumni, S.~Iraoui, K.~Masmar and S.~Zhizeh,
{\hypersetup{urlcolor=vividviolet}\href{https://inspirehep.net/literature/1670197}{LHEP \textbf{02}, 05
(2018)}}, \href{https://arxiv.org/abs/1804.10042}{[arXiv:gr-qc/1804.10042]}.

\bibitem{MahdavianYekta:2019dwf}
D.~Mahdavian Yekta, A.~Hadikhani and \"O.~\"Okc\"u,
``Joule-Thomson expansion of charged AdS black holes in Rainbow gravity'',
{\hypersetup{urlcolor=vividviolet}\href{https://inspirehep.net/literature/1733898}{Phys. Lett. B \textbf{795} (2019), 521-527}},
\href{https://arxiv.org/abs/1905.03057}{[arXiv:hep-th/1905.03057]}.

\bibitem{Mo:2018rgq}
J.~X.~Mo, G.~Q.~Li, S.~Q.~Lan and X.~B.~Xu,
``Joule-Thomson expansion of $d$-dimensional charged AdS black holes'',
{\hypersetup{urlcolor=vividviolet}\href{https://inspirehep.net/literature/1666856}{Phys. Rev. D \textbf{98}, no.12, 124032
(2018)}}, \href{https://arxiv.org/abs/1804.02650}{[arXiv:gr-qc/1804.02650]}.

\bibitem{Kuang:2018goo}
X.~M.~Kuang, B.~Liu and A.~\"Ovg\"un,
{\hypersetup{urlcolor=vividviolet}\href{https://inspirehep.net/literature/1684232}{Eur. Phys. J. C \textbf{78}, no.10, 840
(2018)}}, \href{https://arxiv.org/abs/1807.10447}{[arXiv:gr-qc/1807.10447]};
J.~T.~Xing, Y.~Meng and X.~M.~Kuang,
``Joule-Thomson expansion for hairy black holes'',
{\hypersetup{urlcolor=vividviolet}\href{https://inspirehep.net/literature/1914413}{Phys. Lett. B \textbf{820}, 136604 (2021)}}.

\bibitem{Cisterna:2018jqg}
A.~Cisterna, S.~Q.~Hu and X.~M.~Kuang,
``Joule-Thomson expansion in AdS black holes with momentum relaxation'',
{\hypersetup{urlcolor=vividviolet}\href{https://inspirehep.net/literature/1689404}{Phys. Lett. B \textbf{797}, 134883
(2019)}}, \href{https://arxiv.org/abs/1808.07392}{[arXiv:gr-qc/1808.07392]}.

\bibitem{Liang:2021xny}
J.~Liang, B.~Mu and P.~Wang,
``Joule-Thomson expansion of lower-dimensional black holes'',
{\hypersetup{urlcolor=vividviolet}\href{https://inspirehep.net/literature/1859172}{Phys. Rev. D \textbf{104}, no.12, 124003
(2021)}}, \href{https://arxiv.org/abs/2104.08841}{[arXiv:gr-qc/2104.08841]}.
%

\bibitem{DiValentino:2019qzk}
E.~Di Valentino, A.~Melchiorri and J.~Silk,
``Planck evidence for a closed Universe and a possible crisis for cosmology'',
{\hypersetup{urlcolor=vividviolet}\href{https://inspirehep.net/literature/1762800}{Nature Astron. \textbf{4}, no.2, 196-203 (2019)}}, \href{https://arxiv.org/abs/1911.02087}{[arXiv:astro-ph.CO/1911.02087]}.
%
\bibitem{Handley:2019tkm}
W.~Handley,
``Curvature tension: evidence for a closed universe'',
{\hypersetup{urlcolor=vividviolet}\href{https://inspirehep.net/literature/1751120}{Phys. Rev. D \textbf{103} (2021) no.4, L041301}}, \href{https://arxiv.org/abs/1908.09139}{[arXiv:astro-ph.CO/1908.09139]}.
%
\bibitem{DiDio:2016ykq}
E.~Di Dio, F.~Montanari, A.~Raccanelli, R.~Durrer, M.~Kamionkowski and J.~Lesgourgues,
``Curvature constraints from Large Scale Structure'',
{\hypersetup{urlcolor=vividviolet}\href{https://inspirehep.net/literature/1436471}{JCAP \textbf{06}, 013 (2016)}}, \href{https://arxiv.org/abs/1603.09073}{[arXiv:astro-ph.CO/1603.09073]}.
%
\bibitem{Bel:2022iuf}
J.~Bel, J.~Larena, R.~Maartens, C.~Marinoni and L.~Perenon,
``Constraining spatial curvature with large-scale structure'',
{\hypersetup{urlcolor=vividviolet}\href{https://inspirehep.net/literature/2092637}{JCAP \textbf{09} (2022), 076}},
\href{https://arxiv.org/abs/2206.03059}{[arXiv:astro-ph.CO/2206.03059]}.


\bibitem{Hayward:1994bu}
S.~A.~Hayward,
``Gravitational energy in spherical symmetry'',
{\hypersetup{urlcolor=vividviolet}\href{https://inspirehep.net/literature/375366}{Phys. Rev. D \textbf{53}, 1938-1949 (1996)}}, \href{https://arxiv.org/pdf/gr-qc/9408002}{[arXiv:gr-qc/9408002]}.

\bibitem{Cai:2008gw}
R.~G.~Cai, L.~M.~Cao, and Y.~P.~Hu,
``Hawking Radiation of Apparent Horizon in a FRW Universe'',
{\hypersetup{urlcolor=vividviolet}\href{https://inspirehep.net/literature/796050}{Class. Quant. Grav. \textbf{26}, 155018 (2009)}},
\href{https://arxiv.org/pdf/0809.1554}{[arXiv:hep-th/0809.1554]};
Y.~P.~Hu,
{\hypersetup{urlcolor=vividviolet}\href{https://inspirehep.net/literature/862414}{Phys. Lett. B \textbf{701}, 269-274 (2011)}},
\href{https://arxiv.org/abs/1007.4044}{[arXiv:gr-qc/1007.4044]}.

\bibitem{Dolan:2013ft}
B.~P.~Dolan, D.~Kastor, D.~Kubiznak, R.~B.~Mann, and J.~Traschen,
``Thermodynamic Volumes and Isoperimetric Inequalities for de Sitter Black Holes'',
{\hypersetup{urlcolor=vividviolet}\href{https://inspirehep.net/literature/1216285}{Phys. Rev. D \textbf{87}, no.10, 104017 (2013)}},
\href{https://arxiv.org/pdf/1301.5926.pdf}{[arXiv:hep-th/1301.5926]}.

\bibitem{Zhang:2019oes}
H.~Zhang, Y.~P.~Hu and Y.~Zhang,
``Towards a sound massive cosmology'',
{\hypersetup{urlcolor=vividviolet}\href{https://inspirehep.net/literature/1713364}{Phys. Dark Univ. \textbf{23} (2019), 100257}},
\href{https://arxiv.org/abs/1901.09331}{[arXiv:gr-qc/1901.09331]}.

\bibitem{Banihashemi:2022jys}
B.~Banihashemi and T.~Jacobson,
``Thermodynamic ensembles with cosmological horizons'',
{\hypersetup{urlcolor=vividviolet}\href{https://inspirehep.net/literature/2066030}{JHEP \textbf{07} (2022), 042}}, \href{https://arxiv.org/abs/2204.05324}{[arXiv:hep-th/2204.05324]}.

\bibitem{Banihashemi:2022htw}
B.~Banihashemi, T.~Jacobson, A.~Svesko and M.~Visser,
``The minus sign in the first law of de Sitter horizons'',
\href{https://arxiv.org/abs/2208.11706}{[arXiv:hep-th/2208.11706]}.

\bibitem{Abdusattar:2022bpg}
H.~Abdusattar, S.~B.~Kong, Y.~Yin and Y.~P.~Hu,
``The Hawking-Page-like phase transition from FRW spacetime to McVittie black hole'',
{\hypersetup{urlcolor=vividviolet}\href{https://inspirehep.net/literature/2055705}{JCAP \textbf{08}, no.08, 060 (2022)}}, \href{https://arxiv.org/abs/2203.10868}{[arXiv:gr-qc/2203.10868]}.


\end{thebibliography}
\end{document}